\def\co{{$^{56}{\rm Co}$}}
\def\ej{{\rm ej}}

\def\iso{{\rm iso}}
\def\jet{{\rm jet}}
\def\nk{{$^{56}{\rm Ni}$}}
\def\p{{\rm peak}}
\def\sn{{\rm SN}}

\input{aipcheck.tex}

\newcommand\sps{\space\space\space\space}
\documentclass[
    ,final            % use final for the camera ready runs
  ]
  {aipproc}

\layoutstyle{8x11single}

\SetInternalRegister\hbadness{8000} % pseudo latin isn't breaking very well :-)

\newcommand\doingARLO[2][]{%
  \ifx\mmref\undefined #1\else #2\fi
}

\begin{document}

\title 
      [The GRB-Supernova Connection]
      {The GRB-Supernova Connection}


\author{Li-Xin Li}{
  address={Max-Planck-Institut f\"ur Astrophysik, 85741 Garching, Germany, 
  and\\
  Kavli Institute for Astronomy and Astrophysics, Peking University, 
  Beijing 100871, China},
  email={lxl@mpa-garching.mpg.de, lxl@kiaa.pku.edu.cn},
  %thanks={This work was commissioned by the AIP}
}

% \copyrightholder{Acoustical Scociety of America}
\copyrightyear  {2001}

\begin{abstract}
Long-duration gamma-ray bursts (GRBs) are believed to be produced by the
core collapse of massive stars and hence to be connected with supernovae 
(SNe). Indeed, for four pairs of GRBs and SNe, spectroscopically 
confirmed connection has been firmly established. For more than a dozen of 
GRBs the SN signature (the `red bump') has been detected in the afterglow 
lightcurves. Based on the four pairs of GRBs and SNe with spectroscopically 
confirmed connection a tight correlation was found between the peak spectral
energy of GRBs and the peak bolometric luminosity of the underlying SNe.
The recent discovery of X-ray flash 080109 associated with a normal 
core-collapse SN 2008D confirmed this relation and extended the GRB-SN 
connection. Progress on the GRB-SN connection is briefly reviewed.
\end{abstract}

\date{\today}

\maketitle

\section{Introduction}

%\subsection{\em Gamma-Ray Bursts}

{\em Gamma-ray bursts} (GRBs) are short and intense pulses of soft 
$\gamma$-rays and the brightest objects in the Universe. The key observed 
features of GRBs are as follows (\cite{zha04,mes06} and other review
articles):

\begin{itemize}

\item{The observed durations of GRBs are generally in the range
0.01--1000s.}

\item{The spectra of GRBs are generally nonthermal, spanning a very broad band
from radio to $\gamma$-ray.}

\item{GRBs are characterized by a very large peak luminosity, $10^{50}$--$
10^{53}$ erg s$^{-1}$.}

\item{The total isotropic-equivalent energy emitted by a GRB in 1--10000
keV is $10^{50}$--$10^{54}$ erg, in the extremely energetic case comparable
to the rest mass energy of Sun.}

\item{The distribution of GRBs on the sky is isotropic.}

\item{GRBs are at cosmological distances. At present the highest measured GRB
redshift is $z=6.29$, comparable to that of the remotest quasar.}

\item{The large amount of energy, the non-thermal spectra, and the very
short variability timescale of GRB lightcurves (can be as short as 
milliseconds) indicate that GRBs are relativistic: the outflow producing the 
GRB expands with a Lorentz factor $> 100$, in agreement with direct
measurements \cite{mol07}.}

\end{itemize}

GRBs are the most powerful explosion since the Big Bang. Since they are 
observable to very high redshift ($z>10$), GRBs are very useful for probing 
cosmology.

GRBs are usually classified by their durations: those with durations smaller
than 2 s are called short GRBs, and those with durations larger than 2 s
are called long GRBs. This classification is purely empirical and very 
inaccurate, and sometimes ambiguity in classification may occur. The duration 
is defined in the observer frame and the duration
distribution of the two classes significantly overlaps. Indeed, some GRBs 
with durations greater than 2 s have been argued to be of the same nature 
of short GRBs, e.g. GRBs 060505 and 060614 \cite{zha07,ofe07}.

It is generally thought that long GRBs are produced by the core collapse of
massive stars \cite{woo06}: the iron core of a rapidly spinning massive star 
(main sequence mass $> 30 M_\odot$) collapses to a black hole and an 
accretion disk forms around the black hole. Two oppositely directed jets, 
powered either by the disk accretion or the black hole's spin energy, produce
the observed GRB. This collapsar model is supported by current observations: 
long GRBs are always found in star-forming galaxies, and some are found to be
associated with core-collapse supernovae (SNe) \cite{woo06b}.

In contrast, short GRBs are found in both star-forming and non-star-forming
galaxies and are not associated with SNe \cite{ber06}. They are thought to 
be produced by merger of two neutron stars or merger of a black hole and a 
neutron star.

%\subsection{\em Supernovae}
\vspace{0.3cm}

{\em Supernovae} are stellar explosion at the end of a star's life. 
Electromagnetic emissions resulted from a SN event can last a very long time, 
from several years to centuries. The still bright Crab Nebula is the remnant 
of a SN that exploded about a thousand years ago in the constellation of Taurus 
($\sim 2$ kpc from Earth).

The bulk of the SN electromagnetic emission is in the optical band, with a peak 
luminosity up to $\sim 10^{43}$ erg s$^{-1}$, intrinsically $10^{10}$ times 
brighter than Sun but $10^{10}$ times fainter than bright GRBs.

Most energy released by gravitational collapse of a progenitor star is 
carried away by gravitational waves and neutrinos ($\sim 10^{53}$ erg).  
A small part is converted by shock waves to the kinetic energy of the 
expanding fluid ($\sim 10^{51}$ erg), of which a fraction is emitted as 
electromagnetic radiation ($\sim 10^{49}$ erg) through radioactive decays.

SNe are usually nonrelativistic, with an expansion speed much smaller
than the speed of light. In an extreme case (e.g., SN 1998bw), the expansion
speed can reach $\sim 0.3c$. Because of the low luminosity of SNe relative
to that of GRBs, usually SNe can only be observed to a noncosmological
distance, and SNe discovered at $z>1$ are rare \cite{kuz08}.

In spite of the fact that observations of SNe have a much longer history 
than that of GRBs and hence our understanding of SNe is much better than 
that of GRBs, many important issues of SNe (e.g., the explosion mechanism) 
are still not solved \cite{jan07}. 

SNe are classified by their spectra. SNe having hydrogen lines in their
spectra are called Type II, otherwise called Type I. Type I SNe are further 
divided into three sub-classes. Type I SNe with silicon lines are called 
Type Ia. Type I SNe without silicon lines but with helium lines are called 
Type Ib. Type I SNe without silicon lines and without or with weak helium 
lines are called Type Ic.

SNe Ia, believed to be produced by thermal nuclear explosion of white dwarfs, 
are most luminous and often used as standard candles for measuring cosmological
parameters. SNe Ibc and SNe II are generally thought to be produced by the 
core collapse of massive stars and are called core-collapse SNe. 

\section{The GRB-SN connection}

A surprising discovery in the past decade of GRB observations is that
GRBs and SNe---two seemingly very different phenomena, with very different
time durations, expansion velocities, and photon energy scales---are related.

On 25 April 1998, a faint GRB was detected by BeppoSax and BATSE, which has
a smooth Fast-Rise-Exponentially-Decay (FRED) shape lightcurve with a
duration $\sim 35$ s. About two and half days after the GRB, a bright
SN 1998bw was discovered in the BeppoSax error box of the GRB \cite{gal98}. 
The SN was one of the most unusual Type Ic SNe ever seen. It is very 
bright---comparable to a typical SN Ia---and has very strong radio emissions 
indicating relativistic expansion (the derived expansion speed $\sim 0.3 c$
\cite{kul98}). The small probability for a chance coincidence of GRB 980425 
and SN 1998bw ($\sim 10^{-4}$) indicates that the two events are associated 
\cite{gal98}. 

The host galaxy ESO184-G82 of this GRB has a very low redshift: $z=0.0085$, 
which makes GRB 980425 the nearest GRB so far discovered with redshift. 
With this near distance, GRB 980425 is intrinsically at least $10^4$ times 
fainter than cosmological GRBs.

Since GRB 980425/SN 1998bw, four pairs of GRBs-SNe with spectroscopically 
confirmed connection have been found, which are summarized in Table
\ref{grb_sn} (taken from \cite{li06}).

\begin{table}
\begin{tabular}{lllllllll}
\hline
  \tablehead{1}{l}{b}{GRB/SN}
  & \tablehead{1}{l}{b}{$z$\tablenote{Cosmic redshift.}}
  & \tablehead{1}{l}{b}{$E_{\gamma,\p}$\tablenote{Peak energy of the integrated GRB spectrum in units of keV, measured in the GRB frame.}}
  & \tablehead{1}{l}{b}{$E_{\gamma,\iso}$\tablenote{Isotropic-equivalent energy of the GRB in units of $10^{52}$ erg, defined in the $1$--$10000$ keV energy band in the GRB frame.}}   
  & \tablehead{1}{l}{b}{$M_{\sn,\p}$\tablenote{Peak bolometric magnitude of the supernova, defined in the $3000$--$24000$ \AA\, wavelength band in the SN frame.}}
  & \tablehead{1}{l}{b}{$E_K$\tablenote{Explosion kinetic energy of the
  supernova in units of $10^{52}$ erg.}}
  & \tablehead{1}{l}{b}{$M_\ej$\tablenote{Ejected mass in the supernova explosion in units of $M_\odot$.}}
  & \tablehead{1}{l}{b}{$M_{\rm Nickel}$\tablenote{Mass of \nk\, produced by the SN explosion in units of $M_\odot$.}}   \\
\hline
980425/1998bw & $0.0085$ & $55\pm 21$ & $0.00009\pm 0.00002$ & $-18.65\pm 0.20$
& $5.0\pm 0.5$ & $10\pm 1$ & $0.38$--$0.48$ \\
030329/2003dh & $0.1687$ & $79\pm 3$  & $1.7\pm 0.2$ & $-18.79\pm 0.23$ &
$4.0\pm 1.0$ & $8\pm 2$ & $0.25$--$0.45$ \\
031203/2003lw & $0.1055$ & $159\pm 51$ & $0.009\pm 0.004$ & $-18.92\pm 0.20$ &
$6.0\pm 1.0$ & $13\pm 2$ & $0.45$--$0.65$ \\
060218/2006aj & $0.0335$ & $4.9\pm 0.4$ & $0.0059\pm 0.0003$ & $-18.16\pm 0.20$
& $0.2\pm 0.02$ & $2\pm 0.2$ & $0.2\pm 0.04$ \\
\hline
\end{tabular}
%\source{Central Statistical Office, UK}
\caption{Gamma-ray bursts and supernovae with spectroscopically confirmed
connection ($H_0 = 72$ km s$^{-1}$ Mpc$^{-1}$, $\Omega_m = 0.28$, and
$\Omega_\Lambda = 0.72$; see \cite{li06} for references)}
\label{grb_sn}
\end{table}

All the GRB-SNe are among a special subclass of SNe Ibc: broad-lined SNe,
characterized by relative smooth spectra and a very large explosion
energy. So they are often called `hypernovae'. 

GRB 060218 was detected by Swift. It is the second nearest GRB with redshift. 
The burst is very faint but has an extremely long duration ($\sim 2000$ s). 
A SN associated with it, SN 2006aj, was discovered about three days 
after the GRB trigger. The XRT and UVOT onboard Swift started observing the 
GRB $\sim 156$ s after the trigger, so a very complete multi-wavelength 
observation on the event has been obtained \cite{cam06,pia06,sod06}. 
The discovery of GRB 060218/SN 2006aj with high quality multi-wavelength
spectra and lightcurve data put the GRB-SN connection on a solid foundation.

A remarkable blackbody component was discovered in the X-ray and UV-optical
lightcurve of GRB 060218 (occupying $20\%$ of the tocal emission and lasting 
up to 10000 s in the 0.3--10 keV X-ray emission), which was interpreted 
as SN shock breakout in the progenitor Wolf-Rayet star \cite{cam06}. However,
detailed calculations do not support the shock breakout interpretation
\cite{li07}.

Based on the observation of the four pairs of GRBs and SNe, a strong
correlation between the GRB peak spectral energy ($E_{\gamma,\p}$) and the 
SN maximum bolometric luminosity ($L_{\sn,\p}$; or equivalently, the mass 
of \nk\ generated by the SN explosion, $M_{\rm Nickel}$) was derived 
\cite{li06}
\begin{eqnarray}
	E_{\gamma,\p} = 90.2\,{\rm keV} \left(\frac{L_{\sn,\p}}{10^{43} 
		{\rm erg\,s}^{-1}}\right)^{4.97} \;.
	\label{ep_lp}
\end{eqnarray}
Combination of this relation with the Amati relation \cite{ama06} led to a 
constraint on the isotropic-equivalent energy of a GRB associated with a 
SN with a given maximum luminosity \cite{li06}.

\section{XRF 080109/SN 2008D}

\begin{figure}
\includegraphics[width=9cm]{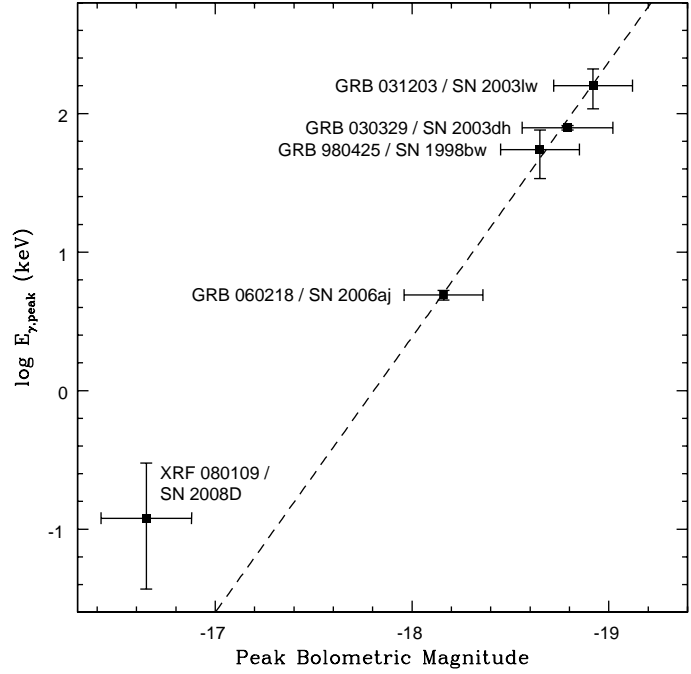}
\caption{The peak spectral energy of GRBs/XRFs versus the peak bolometric 
magnitude of the underlying SNe (from \cite{li08}). The straight dashed line
is the best fit to the four pairs of GRBs/SNe in Table \ref{grb_sn},
$\log E_{\gamma,\p} = -35.38 - 1.987\, M_{\sn,\p}$ (eq.~\ref{ep_lp}).
\label{fig1}}
\end{figure}

\begin{figure}
\includegraphics[width=9cm]{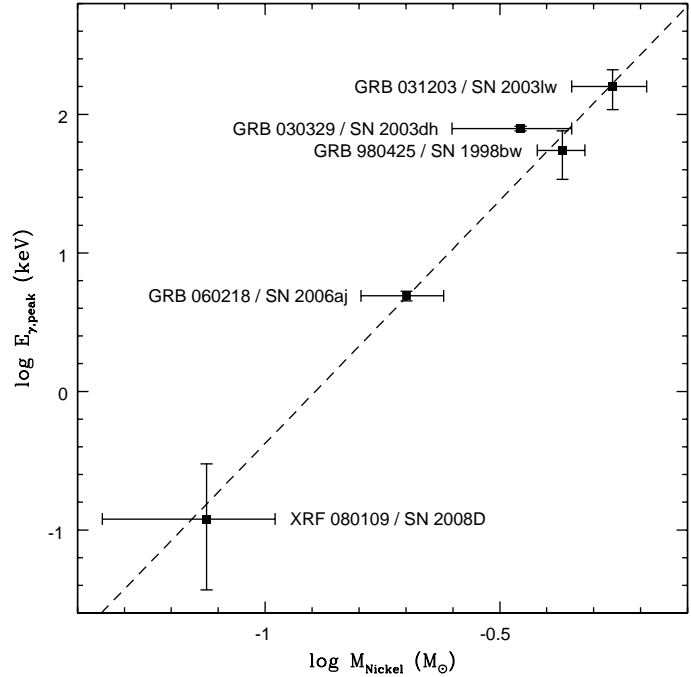}
\caption{The peak spectral energy of GRBs/XRFs versus the mass of
\nk\, generated in the ejecta of the underlying SNe (from \cite{li08}). 
The straight dashed line is the best fit to the four pairs of GRBs/SNe 
in Table \ref{grb_sn}, $\log E_{\gamma,\p} = 3.13 + 3.51 \log M_{\rm Nickel}$.
\label{fig2}}
\end{figure}

The $E_{\gamma,\p}$--$L_{\sn,\p}$ relation in equation (\ref{ep_lp}) was
derived from four `hypernovae' with GRBs. What does it imply if the relation
is applied to a normal Type Ibc SN? This question was answered in \cite{li06}
where it was inferred that ``if normal Type Ibc SNe are accompanied by 
GRBs, the GRBs should be extremely underluminous in the gamma-ray band 
despite their close distances. Their peak spectral energy is expected to 
be in the soft X-ray and UV band, so they may be easier to detect with an 
X-ray or UV detector than with a gamma-ray detector.'' This `prediction'
seems to be confirmed by a recent discovery.

On 9 January 2008, a bright X-ray transient was discovered in spiral
galaxy NGC 2770 during a follow-up observation of SN 2007uy in it by XRT/Swift.
The transient has a FRED shape lightcurve and a duration $\sim 600$ s.
Given the redshift $z=0.006494$ of the galaxy, the average luminosity of 
the transient is $\approx 2\times 10^{43}$ erg s$^{-1}$. At the same position 
of the X-ray transient, a Type Ib SN 2008D was found later \cite{sod08,mod08}.

Although the nature of the transient is debatable, the most natural 
interpretation appears to be an X-ray flash (XRF)---the soft version of
a GRB. Other interpretations (X-ray flare of a GRB, or SN shock breakout)
do not look plausible \cite{xu08,li08,maz08}.

The isotropic equivalent energy derived from an absorbed power-law fit of 
the XRT spectrum is $1.3\times 10^{46}$ erg in 1--10000 keV. A joint analysis 
on the XRT and UVOT data leads to a constraint on the peak spectral energy:
0.037 keV $<E_{\gamma,\p}<$ 0.3 keV. With these results, XRF 080109 satisfies
the Amati relation \cite{li08}.

The peak of the SN lightcurve occurred at about 20~day after the explosion, 
with a peak bolometric magnitude $\approx -16.65$ (corresponding to a maximum 
luminosity $\approx 1.4\times 10^{42}$ erg s$^{-1}$). Fitting the 
lightcurve by an analytic model of SN emission powered by the radioactive 
decay of \nk\, and \co\, yielded a \nk\ mass synthesized in the explosion 
between $0.05$ and $0.1 M_\odot$ \cite{sod08}. These results are in agreement 
with more sophisticated modeling \cite{maz08,tan08}. 

The above results, together with the $E_{\gamma,\p}$ of XRF 080109 derived 
from the XRT and UVOT data, indicate that XRF 080109/SN 2008D agree with the 
$E_{\gamma,\p}$--$L_{\sn,\p}$ relation (Fig. \ref{fig1}) and the 
$E_{\gamma,\p}$--$M_{\rm Nickel}$ relation (Fig. \ref{fig2}) \cite{li08}.

SN 2008D is a normal Type Ibc SN in terms of its luminosity and spectra, in 
contrast to other GRB-hypernovae. The detection of XRF 080109/SN 2008D 
extends the GRB-SN connection to normal core-collapse SNe. It may suggest 
that every Type Ibc (maybe Type II also) SN has a GRB/XRF associated with 
it. If this is true, events like XRF 080109 would occur at a rate comparable 
to that of SNe Ibc, $\sim 10^{-3}$ yr$^{-1}$ in an average galaxy \cite{pod04}.

\section{Spherical GRBs}

In the standard collapsar model of GRBs, collimation of the outflow 
is essential for avoiding baryon loading and producing a clean fireball
with a Lorentz factor $> 100$. However, it appears that some GRBs/XRFs are 
spherical. 

\begin{figure}
\includegraphics[width=9cm]{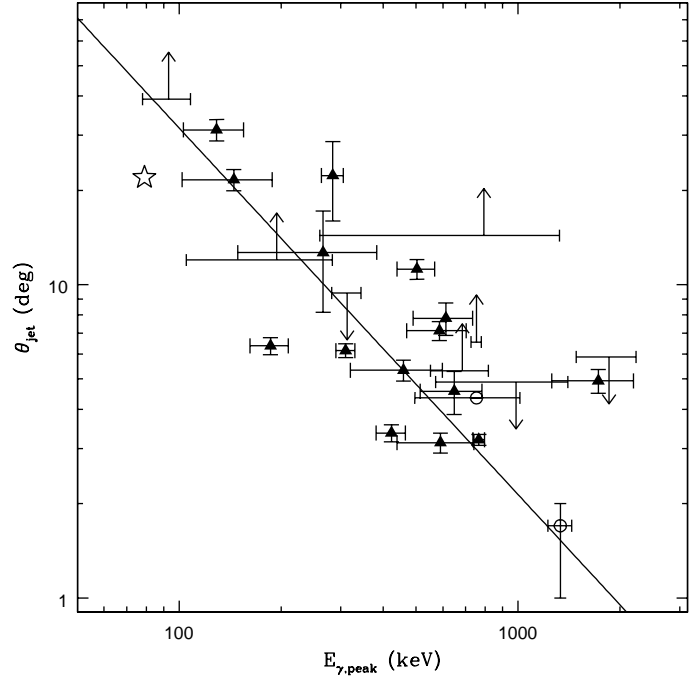}
\caption{The jet opening angle of GRBs versus the peak energy of their 
spectra measured in the GRB frame (from \cite{li06}). The straight line is 
a maximum-likelihood fit to the data excluding the 8 GRBs with only limits,
$\log \theta_\jet = 3.84 - 1.17\, \log E_{\gamma,\p}$.
\label{fig3}}
\end{figure}

Investigations found that GRBs with softer spectra tend to have larger 
jet opening angles i.e. weakly collimated outflows (Fig. \ref{fig3})
\cite{lam05,li06}. For GRBs/XRFs with $E_{\gamma,\p}<40$ keV (in the burst 
frame), the jet opening angle inferred from the above anti-correlation is 
so large that the burst outflow should be spherical \cite{li06}. This is 
consistent with radio observations on the soft XRF 020903, GRB 060218, and 
XRF 080109 \cite{sod04,sod06,sod08}.

Apparently, the standard internal/external-shock fireball model does not apply 
to GRBs/XRFs with a spherical outflow, since a spherical fireball outflow 
cannot avoid baryon loading efficiently: it must pass through the dense
SN ejecta. Then, an unavoidable consequence of a spherical GRB/XRF is that
the outflow producing the burst and the afterglow cannot have a very 
large Lorentz factor. Due to the loss of energy to the SN ejecta, the
burst would also be very sub-energetic compared to normal GRBs. 

Whether a spherical explosion can produce a GRB-like event is a question.
The GRB fireball is trapped inside the heavy SN envelope so the energy
of it may well be dissipated by the SN envelope without producing
a GRB/XRF. However, two possible scenarios for producing a GRB/XRF from a
spherical configuration can be imagined \cite{li08}.

{\em Scenario A.} When a light fluid is accelerated into a heavy fluid, which
is just the case of a spherical GRB explosion as outlined above, the 
Rayleigh-Taylor instability occurs. In a spherical 
GRB/SN explosion, the GRB fireball may emerge from the SN envelope through 
the Rayleigh-Taylor instability, then produce a GRB/XRF through either 
the internal-shock or the external-shock interaction.

{\em Scenario B.} The initial GRB fireball is killed by the SN 
envelope and the fireball energy is added to the SN explosion energy. 
A small fraction of the outer layer of the SN envelope is accelerated 
by the enhanced SN shock wave to a mildly relativistic velocity and 
generates a low-luminosity GRB/XRF via interaction with surrounding matter. 
This GRB-production mechanism through acceleration of the SN outer
layer has been proposed for explaining the prompt emission of GRB 980425
\cite{tan01}. Although this mechanism is able to account for the total
X-ray energy emitted by XRF 080109, it is unable to explain GRB 060218
\cite{li08}.

\section{Summary}

So far, five pairs of nearby GRBs (or XRFs) and SNe with spectroscopically 
confirmed connection have been discovered: the four GRBs/SNe listed in Table 
\ref{grb_sn}, and XRF 080109/SN 2008D. There appears to be a relation 
between the GRB/XRF peak spectral energy and the SN peak luminosity (or the 
Nickel mass). The relation need be tested with future detection of GRB-SN 
and XRF-SN pairs.

Observing SN signatures in high-redshift GRBs is difficult, since by 
selection effects the observable GRBs at high redshift are bright and 
hence the underlying SNe are easily overshone by the GRB afterglows. 
In spite of this challenge, some GRBs have shown rebrightening and 
flattening in their late optical afterglows, which have been interpreted as 
emergence of the underlying SN lightcurve. A systematic study on the GRB 
afterglows with this approach suggests that all long GRBs are associated 
with SNe \cite{zeh04}. 

However, one must be cautious about the above conclusion, since alternative 
explanations for the rebrightening and flattening in the late optical 
afterglows of GRBs exist \cite{esi00,wax00}.

In addition, some nearby long GRBs have not been found to have SNe in spite 
of extensive deep observations, including GRB 060505 and GRB 060614 
\cite{geh06,fyn06,del06}. Failed SNe have
been predicted in theoretical study on the SN explosion and have been
suggested to represent the main mechanism for producing cosmological
GRBs (\cite{woo93,gou02}, see however, \cite{mac99}). Hence, it is possible
that not every long GRB is associated with a SN.

On the other hand, although the discovery of XRF 080109 with SN 2008D may
indicate that every SN Ibc has a preceding GRB or XRF-like event, it may 
also be true that not all core-collapse SNe are associated with GRBs/XRFs. 
For example, the broad-lined Type Ic SN 2003jd is only slightly less luminous 
than SN 1998bw but no GRB/XRF has been found to be with it 
\cite{maz05,sod06b}.

Searching for more pairs of GRBs (XRFs) and SNe in future observations is 
very important for understanding the nature of the GRB-SN connection, the
nature of GRBs, and the explosion mechanism of core-collapse SNe.

% choose bibtex style depending on layout style and options used in
% sample:

\doingARLO[\bibliographystyle{aipproc}]
          {\ifthenelse{\equal{\AIPcitestyleselect}{num}}
             {\bibliographystyle{arlonum}}
             {\bibliographystyle{arlobib}}
          }
\bibliography{sample}

\end{document}